\documentclass[11pt]{article}
\def\a{\alpha}

\def\g{\gamma}
\def\d{\delta}

\def\f{\phi}

\def\y{\eta}
\def\x{\xi}
\def\k{\kappa}
\def\l{\lambda}
\def\m{\mu}
\def\n{\nu}
\def\r{\rho}
\def\p{\pi}

\def\x{\xi}
\def\v{\varphi}

\def\D{\Delta}

\def\part{\partial}

\def\Q{ {}^*Q}

\def\be{\begin{equation}}
\def\ee{\end{equation}}
\def\ol{\overline}

\begin{document}
\begin{titlepage}
\noindent
S\~ao Paulo\\
July, 1997\\
IFUSP/P-1265\\
hep-th/9710224\\           % \hskip 8cm {\bf \today}

\vspace*{5 mm}
\vspace*{35mm}
\begin{center}{\LARGE{\bf{ BRST-co-BRST Quantization of Reparametrization
Invariant Theories}}}
\end{center}
\vspace*{3 mm} \begin{center} \vspace*{3 mm}
\begin{center}G\'eza F\"ul\"op\footnote{E-mail: fulop@fma.if.usp.br}\\
%Dmitry M. Gitman \footnote{E-mail: gitman@fma.if.usp.br}\\
\vspace*{7 mm} {\sl Institute of
Mathematical Physics\\ USP\\
Brazil}\end{center}
\vspace*{25 mm}

\begin{abstract}
We study some reparametrization invariant theories 
in context of the BRST-co-BRST quantization method.
The method imposes restrictions on the possible
gauge fixing conditions and leads to well defined
inner product states through a gauge regularisation
procedure. Two explicit examples are also treated in 
detail.
\end{abstract} 
\end{center}
\end{titlepage}

\setcounter{page}{1}
\setcounter{equation}{0}

%{\bf \today}

\section{Introduction}                       \label{bevezetes}
There exists a large class of theories
that are invariant under reparametrization.
In order to quantize these theories one has
to introduce a gauge fixing that relates the initial
time-like variable of the theory to an internal parameter.
This parameter is understood as the physical time of the system.
It means that we obtain a time dependent gauge
fixing condition. The standard Dirac quantization
procedure \cite{dirac} has to be modified for this
case. There are several methods developed to handle
time dependent constraints. 
One of the methods by Gitman and Tyutin
\cite{g-t} decomposes
time evolution into two quite natural components.
Part of time evolution in this treatment
is generated by the Hamiltonian as is the case for
the ordinary theories with time independent constraints.
The other part    
is given by the time variation of the constraint surface.
This part might not lead to a unitary time evolution in
the quantum theory. However, in cases when it does,
one can find an effective Hamiltonian defined on the
physical phase space that generates the same time evolution
for the physical variables as the one found by the 
above mentioned treatment. For some details of this
method see \cite{g-g} and \cite{f-g}.

A purely geometrical method of quantization of time dependent
constraints also exists \cite{evans1} and \cite{evans2}.
This method has the advantage to be able to find possible
topological obstructions in the way of constructing
a globally defined effective Hamiltonian. 
It does not give however a recipe on how to find the
physical variables of the theory.

There also exists a completely different view on this 
issue, presented in \cite{henneaux2}.
It is shown there that one can use the standard
canonical quantization if one modifies the action 
of reparametrization invariant theories 
by adding apropraiate surface terms.

Tese methods are generalizations of the Dirac quantization
procedure. We have however, another very powerful quantization
method, namely BRST-quantization \cite{brst}.
This method has the advantage that together with a co-BRST
operator it leads to a well defined inner product space \cite{kalau},
\cite{f-m}, \cite{fdok}.
The BRST-quantization of theories with time-dependent
constraints and Hamiltonian was done in \cite{batalin}.

The BRST-charge $Q$ (or operator in the quantum case)
is a conserved nil-potent quantity,  built on a linear combination
of the first class constraints of the theory:
\be
Q = \v_i \y^i + ...
\ee
where $\v_i$ are the first class constraints and the
$\y^i$ are the so called ghost variables, which have different
Grassmann
parity from the constraints: if a constraint is Grassmann-even,
the corresponding ghost is odd, and vice-versa.

Co-BRST $\Q$ is built on the combination of the gauge fixing
conditions in a way quite similar to the one used in the
case of the BRST charge.
In our case, where  the gauge fixing is necessarily 
time dependent, co-BRST will depend on time too: 
it will contain the information on how the physical time $t$ is related
to one of the configuration space variables ($x^0$).
%the internal parameter $t$.
We shall see that there are some conditions that arise from
the demand that there exist physical states (or functions in the
classical case) in the theory.

There exists an important difference between reparametrization
invariant theories and other models with time dependent
constraints.
In the case of the latter ones the effective Hamiltonian 
generates the physical time evolution. 
In our case however, we may choose two different
interpretations.
In one of them $t$ is  the
physical time and the effective Hamiltonian, being the physical
Hamiltonian, generates
the evolution along it.
In the other interpretation however, $t$ is only an internal 
parameter and the physical Hamiltonian has to be found separately
and it generates evolution along $x^0$.
It will then only depend on the physical variables.

The second interpretation is the natural one, when one has an
externally given  time.
However, for generic reparametrization invariant theories
(like gravitation) the time variable is not a priori given.
For these models the first interpretation is valid.
This is also the interpretation we follow in this paper.
For a more thorough discussion see \cite{henneaux}.
 
There are  many interesting reparametrization invariant models.
As shown in \cite{f-g}, all models with vanishing Hamiltonian
are reparametrization invariant.
At the same time, any other model (both with regular and singular
Lagrangians) can be transformed into a reparametrization invariant
form by making the time variable depend on a parameter
$x^0 \rightarrow x^0(t)$:
\be
S = \int L(q, \frac{dq}{dx^0}, x^0) dx^0
\rightarrow S' = \int L\left(q, \frac{\dot{q}}{\dot{x^0}}, 
x^0\right) \dot{x^0}dt~,                                  \label{rep}
\ee
where $\dot{x^0} = \frac{dx^0}{dt} > 0, \forall t$. 
The action $S'$ is invariant under
the reparametrization of $x^0$.
This transformation of theories into a reparametrization invariant
form can be used to compute the time-evolution in
non-reparametrization
invariant models \cite{rob.time}.

The plan of the paper is the following: In the next Section we 
introduce the two important charges (operators) 
we are going to use later: the BRST and co-BRST quantization.
The in Sections 3 we show the BRST-co-BRST quantization of
a reparametrization invariant theory.
In Sections 4 and 5 we treat two simple  examples,
those of the non-relativistic and the relativistic particle,
written in a reparametrization invariant form.
We show here how the restrictions on the possible gauge-fixing
conditions appear as a consequence of demanding the existence
of a consistent co-BRST charge.
We conclude the paper with a discussion in the last Section.

Throughout the paper we use the following conventions:
in the classical case all variables (even the ghost ones)
come in canonical pairs:
\be
\{x, p \} = 1~, \hskip 1cm \{ \y, {\cal P} \} = 1~.
\ee
In the quantum case we have:
\be
[x,p]=i \hskip 1cm [\y, {\cal P} ] = i, \hskip 1cm \y^{\dagger} = \y,
\hskip 1cm
{\cal P}^{\dagger} = - {\cal P}
\ee

\setcounter{equation}{0}
\section{BRST and co-BRST}

BRST-quantization is based on a symmetry of the 
gauge-fixed Lagrangian \cite{brst}.
In the Hamiltonian picture this is expressed
in the definition of the physical states, namely,
that the physical states are those that get annihilated
by the BRST-operator $Q$.
However, the existence of this symmetry means that there is
freedom left in the states.
The freedom connected to the BRST-symmetry is manifested
through the nil-potency of the BRST operator ($Q^2 =0$),
which is, in some sense, its most important property.
Nil-potency means that all BRST-exact states are also BRST-invariant:
\be
Q (Q \mid i \rangle) = 0, \hskip 4mm \forall  \mid i \rangle~.
\ee
This means that our previous definition of the physical states
is too vague: it contains too many states.
The obvious definition of a physical states should be the
following: {\em physical states} are the representatives of
the cohomology classes of the BRST operator.
There is of course some freedom in choosing a representative,
the same way, as there exists freedom in choosing a
gauge condition.
In simpler words: states that are annihilated by the
BRST operator, but that can not be produced by $Q$ acting
on any other states, should be considered the physical ones:
\be
Q \mid ph \rangle =0, \hskip 5mm \mid ph \rangle \neq Q \mid i
\rangle~.
\ee
The question is then, how to find the cohomology classes
of $Q$. The method followed in this paper is to find an operator,
that can fix the BRST-freedom.
This is done by the so called co-BRST operator.
Its definition was given in \cite{kalau}.
The whole BRST-co-BRST structure resembles closely the 
Hodge structure know in geometry. Using that language
we can say, that we are looking for the BRST-harmonic
states.
The explicit form of the co-BRST operator can be found as
in \cite{f-m}. 
The idea is based on the fact that the state space is not entirely
positive definite: half of the states are
of negative norm. We can then find those metric operators
that ``turn'' these states too into positive definite norm states.
The metric operators ($\hat{\g}$) found this way are used to
produce the co-BRST operator:
\be
\Q = \g Q \g~.                                  \label{kobrst}
\ee  
The previous definition of the physical states is identical to the
following:
$\mid s \rangle$ is a physical state if it satisfies the following
equations:
\be
Q \mid s \rangle = \Q \mid s \rangle = 0~.          \label{fiz}
\ee
What is the advantage of this quantization scheme as compared
to other methods?
A first thing is its elegance: the common solutions of two 
equations contain all the physics of the model.
A more important feature of this method is, that it automatically
leads to well defined inner-product states.
The darker side of the coin is that it is not always entirely clear
what should be the exact for of the gauge fixing
conditions that enter in $\Q$.
The method gives some clue, though, through an $SL(2,R)$
algebra existing between the terms of $\Q$.
This algebra exists on the operator level in the
case of Abelian models, but only in an effective way for
non-Abelian models, when using the simplest possible
forms of the gauge fixing conditions.

The BRST-co-BRST method can also give some information
on the time evolution of the system (see e.g. \cite{en}).
However, there are many details yet to be understood.

Co-BRST, as well as BRST, is also well defined 
on the classical level, as a function on the phase space.
They can even give some information about what kind of
gauge fixing conditions we might use: we have to demand that
the formalism is such that both $Q$ and $\Q$ are conserved
in time. To show how this works, we shall treat two simple classical
examples in the Sections 4 and 5.

\setcounter{equation}{0}
\section{BRST-co-BRST Quantization of a Reparametrization Invariant Model}

Let us consider a regular Lagrangian $L(q,\dot{q})$, where
$q$ represents all the configuration space variables. 
To make this theory reparametrization invariant
we change the variables $t \rightarrow x^0$ and we make
the new variable $x^0$ depend on a parameter $t$.
The new action has the form shown in (\ref{rep}).
The only constraint is
\be
\v_2 = p_0 + H =0~,                                   \label{kenyszer2}
\ee
where $H$ is the Hamiltonian of the original non-reparametrization
invariant theory.
The total  Hamiltonian is
\be
H_t = \l \v_2 =0~.
\ee
The standard form of the phase-space Lagrangian can be written
as:
\be
L = p \dot{q} + p_0 \dot{q}^0 - \l \v~.
\ee
Now we have a primary constraint, the momentum conjugated
to $\l$ being constrained to vanish:
\be
\v_1 \equiv \p_{\l} =0~.
\ee
The secondary constraint is just the one given by (\ref{kenyszer2}).
Both are first class constraints and their algebra is abelian,
meaning that the structure of the BRST-charge will be
very simple: just a linear combination of the constraints,
where the coefficients are the ghost variables.
It also means that in the {\em classical} case BRST-invariance of a
function is equivalent to its vanishing Poisson brackets with the
constraints.
Thus the physical variables in the BRST formulation have to
obey the same conditions as in the canonical formulation.
We expect that the physical configuration space variables (denoted 
from now on by $y$)
should be combinations of the initial variables, containing $q$.
The simplest way to find the physical variables is to
define a classical operator $\hat{O}$, whose action on a function
is the same as taking the Poisson bracket between the Hamiltonian
and that function:
\be
\hat{O} q = \{ H, q \}~.
\ee
Then the Poisson bracket of a physical variables with the
constraint $\v_2$ becomes:
\be
\part_0 + \hat{O}y = 0~, \hskip 5mm {\mbox{where}} 
\hskip 5mm \part_0 \equiv \frac{\part}{\part x^0}.
\ee
The solution to this equation is 
\be
y = e^{x^0 \hat{O}} x~.
\ee
We have to remember that $\hat{O}$ is purely classical
and has nothing to do with the quantum theory.
The momenta conjugated to $y$ is
\be
\p =  e^{x^0 \hat{O}} p~.
\ee
To describe physics, we need gauge fixing conditions for both
first class constraints.
The natural form of these conditions is:
\be
\psi_1 = \l - g(t) =0~, \hskip 5mm
\psi_2 = x^0 - f(t;q,p,p_0) = 0~,
\ee
where the function $f$ can depend on all variables
except $x^0$. As proven in Appendix A of \cite{g-t},
it is always allowed to make such a choice.
We know that the co-BRST charge contains the gauge fixing
conditions.
Our definition of the physical variables demands
that their Poisson brackets with the co-BRST vanish.
Because of the simple structure of $\Q$ this means that 
the Poisson brackets with the gauge fixing conditions
have to vanish.
In two examples (the non-relativistic and the relativistic
particle) we prove that
the gauge-fixing function $f$ can depend in those cases
on the phase space variables only through the 
reparametrization constraint.

The BRST-co-BRST quantization of this model %in the previous Section
can be done along the line followed in \cite{f-m}.
The idea is to find first two simple BRST-invariant states,
which can be written as a direct product of a ghost and a
matter state. These states are not going to be well defined
inner product states: the inner product between a such of a state
with itself is typically $\infty 0$.
We can then apply a gauge regulator that leads us to inner product
states, which is also co-BRST invariant.

In this Section we use a coordinate representation:
\be
\Psi(q,x^0,\l,\y,\ol{\y}) = \langle q,x^0,\l \mid 
\langle \y \ol{\y} \mid \Psi \rangle~.
\ee
 The operators corresponding to the variables in the
classical theory are defined in the standard way:
the coordinates are multiplicative, while the momenta
are derivative operators, e.g.:
\be
p_0 = i \frac{\part}{\part x^0}~,
\hskip 5mm 
{\cal P} = i \frac{\part}{\part \y}~,
\hskip 5mm {\mbox etc.}                            \label{repres}
\ee
The BRST  operator is
\be
Q = -i \p_{\l} \ol{\cal P} + (p_0 + H) \y~.        \label{brst.q}
\ee
We look for  BRST-invariant states of the form 
\be
\mid \v_i \rangle = \mid M_i \rangle \mid Gh_i \rangle,
\ee
where $\mid M_i \rangle$ is a purely matter state, while
$\mid Gh_i \rangle$ is a purely ghost state.
Two simple solutions of this form can be found, 
$\mid \v_i \rangle,~i=1,2$, 
which satisfy the following equations:
\be
\p_{\l} \mid \v_1 \rangle = \y \mid \v_1 \rangle =0 \hskip 5mm
\Rightarrow \hskip 5mm
 \p_{\l} \mid M_1 \rangle = \y \mid Gh_1 \rangle =0~,
\ee
\be
(p_0 + H) \mid \v_2 \rangle = \ol{\cal P} \mid \v_2 \rangle =0 \hskip 5mm
\Rightarrow \hskip 5mm
(p_0 + H) \mid M_2 \rangle = \ol{\cal P} \mid Gh_2 \rangle =0~,
\ee
We know from \cite{rob.antib}, that one can always impose
some conditions on the BRST-invariant states without changing
their physical content.
One of the possible restrictions is the demand that the states
should also be anti-BRST invariant.
The anti-BRST operator in this model is
\be
\ol{Q} = i \p_{\l} {\cal P} + (p+H_0) \ol{\y}.
\ee 
Anti-BRST invariance means the following conditions on our states:
\be
 \ol{\y} \mid \v_1 \rangle =0 \hskip 5mm
\Rightarrow \hskip 5mm
 \ol{\y} \mid Gh_1 \rangle =0~,
\ee
\be
{\cal P} \mid \v_2 \rangle =0 \hskip 5mm
\Rightarrow \hskip 5mm
{\cal P} \mid Gh_2 \rangle =0~.
\ee
There is one more freedom left in both states, which can be
fixed by a gauge fixing condition:
\be
(x^0 - f(t)) \mid \v_1 \rangle =(x^0 - f(t)) \mid M_1 \rangle =0~,
\ee
\be
( \l - g) \mid \v_2 \rangle = (\l -g) \mid M_2 \rangle =0~.
\ee
The problem with these two states is that they are not well defined
inner product states either. (And as we shall see later,
they are not co-BRST invariant either.)
As it can be controlled easily, the inner product of a state 
$  \mid \v_i \rangle$ with itself is not a finite number:
\be
\langle \v_i \mid \v_i \rangle =  \langle M_i \mid M_i \rangle 
\langle Gh_i \mid Gh_i \rangle = \infty 0~,~~~~ i = 1,2.
\ee
The inner product between a ghost state and itself vanishes 
because of its Grassmann parity:
\be
\langle \y \mid \y' \rangle = \d(\y - \y') = i (\y - \y')~.
\ee
Following the recipe given in \cite{f-m}, one can define 
the gauge regulators, that take care of both problems related
to the $\mid \v_i \rangle$ states.
Let us define the two singlet states in the following way:
\be
\mid s_i \rangle = e^{\a K_i} \mid \v_i \rangle~, ~~~i=1,2,
\ee
where 
\be 
K_i = [ Q, \r_i]~, ~~~ i =1,2,
\ee
and the operators $\r_i$ are the following
\be
\r_1 = i (\l - g) {\cal P}~, \hskip 1cm \r_2 = (x^0 - f(t)) \ol{\y}.
\ee
This means that 
\be
K_1 = - (\l - g) (p_0 + H) - i \ol{\cal P}{\cal P}~,
\ee
\be
K_2 =  \p_{\l} (x^0 - f(t)) - i \y \ol{\y}~.
\ee
One can prove that the two $K_i$ operators and their commutator
satisfy an $SL(2, {\bf R})$ algebra:
\be
[K_1, K_2] =  K_3, \hskip 5mm
[K_1, K_3] =  -2 K_1, \hskip 5mm
[K_1, K_2] = 2 K_3~.% \hskip 5mm
\ee
The linear combinations of these generators
$X_1 = \frac{1}{2}(K_1+K_2), X_2 = \frac{i}{2}(K_1-K_2), 
X_3 = \frac{i}{2}K_3$ obey the usual 
commutation rules for $SL(2,{\bf R})$. 

As one defined the $\mid \v_i \rangle$ states
by the operators annihilating them, one
can do the same thing here, and define the singlet
states by their annihilating operators.
Denote the operators annihilating $\mid \v_i \rangle$
by $B_{ia}, a= 1,4$. Then the operators that annihilate
$\mid s_i \rangle$ are:
\be
D_{ia} = e^{\a_i K_i} B_{ia} e^{-\a_i K_i}~,
\ee
where $\a_i,~i=1,2,$ are real, finite, non-zero constants.
The operators $D_{ia}$ can be found using
\be
e^A B e^{-A} = B + [A,B] + \frac{1}{2!} [A,[A,B]]+
\frac{1}{3!} [A,[A,[A,B]]]+ ...~.
\ee
This way we find
\[
(\p_{\l} - i \a_1 (p_0 +H)) \mid s_1 \rangle=
(x^0 -f(t) + i \a_1 (\l - g))\mid s_1 \rangle=
\]
\be
(\y + \a_1 \ol{\cal P}) \mid s_1 \rangle=
(\ol{\y} - \a_1 {\cal P}) \mid s_1 \rangle= 0~,          \label{D1}
\ee
\[
(p_0 +H +i \a_2 \p_{\l}) \mid s_2 \rangle=
((\l - g) -i \a_2(x^0 -f(t)) )\mid s_2 \rangle=
\]
\be
( \ol{\cal P}+ \a_2 \y ) \mid s_2 \rangle=
({\cal P} - \a_2 \ol{\y}) \mid s_2 \rangle= 0~.
\ee  
The same set of states $\mid s_1 \rangle$
are be obtained both ways, as can be seen by setting
$\a_1 \a_2 =1$.This means that there is no reason to talk about two
different sets of singlet states. There exists one set
of singlets, whose members can be obtained from
both initial BRST-invariant states $\mid \v_i \rangle$.
For this reason we will drop  the indices $i$ from the 
states and the corresponding constants $\a$.

There exists a relatively simple form of the
BRST and the co-BRST operators based on the 
operators annihilating the singlets (\ref{D1})
or on some linear combinations of them that lead to
a Fock-space representation. The advantage of this
latter representation is that it makes easier
to find the metric operator leading to the explicit form
of the co-BRST operator.
Let us first introduce a new notation for the operators
in (\ref{D1}):
\be
\f = \a (p_0 + H) + i \p_{\l}, \hskip 5mm \x = \frac12 (\l - g) 
- \frac{i}{2\a}(x^0-f(t))~,
\ee
\be
\r = \y + \a \ol{\cal P}~, \hskip 5mm 
\k = \frac{i}{2 \a} \ol{\y} - \frac{i}{2} {\cal P}~.
\ee
The normalization is chosen here to give the commutation relations:
\be
[ \f, \x^{\dagger}] = 1, \hskip 5mm [\k, \r^{\dagger}] =1~,
\ee
the other ones being zero.
Using these operators, the BRST has the following form:
\be
Q = \frac{1}{\a} \left[ \f \r^{\dagger} + \f^{\dagger} \r \right]~.
\ee 
One can now go over to the Fock representation
by defining the following operators:
\be 
a = \x + \frac{1}{2} \f~, \hskip 5mm b = \x - \frac12 \f,
\hskip 5mm
A = \r + \frac12 \k~, \hskip 5mm B = \r - \frac12 \k~.       \label{abAB}
\ee
The non-vanishing commutators between these operators are:
\be
[a, a^{\dagger}] = 1~, \hskip 5mm
[b, b^{\dagger}] = -1~, \hskip 5mm
[A, A^{\dagger}] = 1~, \hskip 5mm
[B, B^{\dagger}] = -1~.
\ee
These commutation relations show, that the operators
$ a, b, A, B; a^{\dagger}, b^{\dagger},  A^{\dagger}, B^{\dagger} $ 
can be regarded as annihilation resp. creation operators.
Half of them ($b, B$) correspond to states with negative 
definite norms. (We have to note here, that the definition
of the operators $a, b, A, B$ can be made more general by
using complex numbers instead of $\frac12$ in (\ref{abAB}).
For a discussion about the general case see \cite{f-m}.)
Now we can define the metric operators as the ones, that change
all states created by $a^{\dagger}, b^{\dagger}, A^{\dagger}, B^{\dagger}$
into positive normed states:
\be
\hat{\g} = e^{i \p b^{\dagger}b} e^{-i \p B^{\dagger} B}~.
\ee
Through the relation given in \cite{kalau} $\Q = \hat{\g} Q \hat{\g}$
we find:
\be
Q = \frac{1}{2\a} \left[ (a-b)(A^{\dagger} + B^{\dagger}) +
(a^{\dagger} - b^{\dagger})(A + B) \right]~, 
\ee
\[
\Q = \frac{1}{2\a} \left[ (a + b)(A^{\dagger} - B^{\dagger}) +
(a^{\dagger} + b^{\dagger})(A - B) \right]~.
\]
Returning to the previous notation the co-BRST operator has the form:
\be
\Q = \frac{1}{\a} \left[ \x \k^{\dagger} + \x^{\dagger} \k \right]
= -\frac14 \left[ i (\l - g) {\cal P} + 
\frac{1}{\a^2} \ol{\y} (x^0 - f(t))\right] ~.
\ee
Now we can write the condition for the physical states in the
simplest possible form. Physical states are those that are annihilated
by both the BRST and the co-BRST operators (\ref{fiz}), or
equivalently
\be
\f \mid s \rangle = \x \mid s \rangle = \r \mid s \rangle = 
\k \mid s \rangle = 0~.                           \label{negy}
\ee
It is clear from here, that in this case the ghost part and
the matter part of the states can be separated: $\mid s \rangle
= \mid s_M \rangle \mid s_{Gh} \rangle$.
%The ghost part becomes:
%\be
%\langle \y, \ol{\y} \mid \Psi \rangle \equiv 
%\Psi_G(\y, \ol{\y}) = e^{ \frac{i}{\a} \ol{\y} \y}~.
%\ee 
To obtain the matter part one has to allow the eigenvalues
of one of the operators become imaginary.
It was shown by Pauli \cite{pauli} that
hermitian operators in indefinite metric
spaces must have imaginary eigenvalues.
Now the solution of the matter part of (\ref{nagy})
can be written as:
\be
\Psi_M(\l, i x^0) \equiv \langle \l \mid
\langle i x^0 \mid s_M \rangle = \d(\l - \frac{1}{\a}(x^0 -f(t)))~,
\ee
where the vectors $\mid i x^0 \rangle$ have the following
properties:
\be
x^0 \mid i x^0 \rangle = i x^0 \mid i x^0 \rangle~, \hskip 5mm
\int dx^0 \mid i x^0 \rangle \langle x^0 \mid = 1~,
\hskip 5mm
\mid i x^0 \rangle^{\dagger} = \langle -ix \mid~.
\ee
The norm of the $ \Psi_M $ state is $\frac{2}{\a}$.
The case of the ghost part is slightly more
complicated. Since $\a$ is real, the eigenstates of $\r$ and
$\k$ corresponding to zero eigenvalue are of the form
$e^{\frac{i}{\a} \ol{\y} \y}$ which are of an imaginary norm.
The only way to change this without altering the commutators
between the ghost operators and their momenta is to
use a ghost-eigenstate  with  an imaginary eigenvalue. 
That means that we change one of the ghost equations in
(\ref{repres}) into:
\be
\hat{\y} \mid i \y \rangle = i \y \mid i \y \rangle~, \hskip 5mm
\hat{\cal P} \mid i \y \rangle = \frac{\part}{\part \y} \mid i \y
\rangle ~.
\ee 
The equation for the anti-ghost remains the same.
The ghost states then have the norm: $\frac{\a}{2}$, meaning
that the singlets have a unit norm.

>From the point of view of the physical states, the singlets
defined before correspond to the vacuum. 
The other physical states have to be defined by
acting with physical creation operators on $\mid s \rangle$.
These operators have to commute both with the BRST and the co-BRST
operator and are typically of the form $p \pm iq$.
%co-BRST invariance (with the co-BRST operator defined in
%(\ref{cobrst.q})) demands that %$\a_1^2 = \a_2^2 =1$. 

There is one more thing we can notice about this formalism:
a possible relation between the gauge regularisation
and time evolution.
The physical states are supposed to have a time evolution
generated by the effective Hamiltonian:
\be
\mid s(t) \rangle = e^{- i \int (p_0 + H) \part_t f dt}
\mid s(t=0) \rangle~.
\ee
This is astonishingly similar to the definition of the
singlet
\be
\mid s(\a) \rangle = e^{-(p_0 + H) \l \a} \left(
e^{ - i \a {\ol{\cal P}} {\cal P}} \mid \v_1 \rangle \right)~,
\ee
especially, if we remember the original definition of
$\l = \part_t x^0 \approx \part_t f$.
In the formalism we used, we can not set $\a = i t$ since $\a$ is
real. However, in an imaginary time formalism, $\a$ also plays the
role of the evolution parameter.

\setcounter{equation}{0}
\section{The 1-Dimensional Particle in a Potential}

As a first example let us
consider the Lagrangian of a particle in 1+1 dimensions
and make it reparametrization invariant:
\be
\int dt L(x, \dot{x}) \rightarrow 
\int dt L(x, \frac{\dot{x}}{\dot{x}^0})\dot{x}^0~.
\ee
We follow now the procedure in the previous Section
with the explicit form of the constraint $\v_2  = p_0 + H =0, \hskip 5mm 
H = \frac{p^2}{2m} + V(x)$.
The natural form of the gauge  conditions is:
\be
\psi_1 = \l =0~, \hskip 5mm
\psi_2 = x^0 - f(t;x,p,p_0) = 0~,
\ee
where the function $f$ can depend on all variables
except $x^0$.
The non-vanishing part of the constraint algebra is
\be
\{ \v_1, \psi_1 \} = -1~, \hskip 5mm
\{ \v_2 , \psi_2 \} =   -1 + \{f(t), H \}~.    \label{kenyszalg4}
\ee
The BRST charge and the co-BRST charge are of the following form:
\be
Q = \v_2 \y - i \v_1 \ol{\cal P}~, \hskip 5mm
\Q =  i \psi_1 {\cal{P}} + \psi_2 \ol{\y}~.
\ee
The factors $i$ appear to ensure that both $Q$ and $\Q$
are real in the classical and Hermitian in the quantum case.
The Poisson bracket between them:
\be
\{Q, \Q \} =  i \l (p_0 + H) - i \p_{\l} \left[ x^0 - f(t;x,p,p_0)\right]
+ \y \ol\y[ -1 + \{f(t), H \}] - \ol{\cal{P}} {\cal{P}}
\ee
The time evolution of the variables can be computed following
the Gitman-Tyutin scheme \cite{g-t}, which is, as mentioned 
before, a generalization of the Dirac-bracket for the case of
time dependent constraints.
The total time derivative of a variable $q$ is given by:
\be
\frac{d}{dt} q = \{ q, H_t \}_D - 
\{ q, \v_i \} [ \{\v, \v\}^{-1}]^{ij} \frac{\partial \v_j}{\partial t}
\ee
where we denoted all constraints (the gauge fixing conditions
included)
 by $\v_i, i=1,4$. The gauge fixing conditions are denoted here
as $\psi_1 \equiv \v_3, \psi_2 \equiv \v_4$.
The time evolution of our basic variables is then:
\begin{eqnarray}
\frac{d}{dt}x &=&  \frac{p}{m} \frac{\part_t f}{1 - \{f, H \} }~, \label{ido1}
\\
\frac{d}{dt}x^0 &=&  \frac{\part_t f}{1 - \{f, H \} }~, \label{ido2}
\\
\frac{d}{dt} p &=& -\part_x V(x) 
\frac{\part_t f}{1 - \{f, H \} }~, \label{ido3}
\\
\frac{d}{dt} p_0 &=& 0~.                   \label{ido4}
\end{eqnarray}
We see that both  $Q$ and $\Q$ are conserved in $t$.
The physical variables are the ones in the BRST-cohomology,
thus those that have vanishing Poisson brackets both with
the BRST and the co-BRST charge.
We see that neither $x$ nor $p$  satisfy
these conditions. 
\begin{eqnarray}
\{ x, Q \} &=& \frac{p}{m} \y, \hskip 15mm 
\{ x, \Q \} = - \{x, f \} \ol{\y}~,
\\
\{ x^0, Q \} &=&  \y, \hskip 19mm 
\{ x^0, \Q \} = - \{x^0, f \} \ol{\y}~,
\\
\{ p, Q \} &=&  -\part_x V(x) \y~, \hskip 3mm 
\{ p, \Q \} = - \{ p, f \} \ol{\y}~,
\\
\{ p_0, Q \} &=&  0, \hskip 19mm 
\{ p_0, \Q \} = -  \ol{\y}~.
\end{eqnarray}
We can however find some combinations
of the variables we have, to obtain the physical ones.
We would like to obtain physical variables
that are canonically conjugated to each other and that 
do not depend on the ghost variables.
In the light of the previous equations we see
that the BRST-invariance of a variable $y(x,x^0,p,p_0)$ can be
written as:
\be
\part_0 y = \hat{O} y~,   \label{nagy}
\ee where
\be
\part_0 \equiv \frac{\part}{\part x^0} \hskip 5mm \mbox{and}\hskip 5mm
\hat{O} \equiv \frac{dV}{dx} \frac{\part}{\part p} - 
\frac{p}{m}\frac{\part}{\part x}~.
\ee
The effect of operator $\hat{O}$ is identical 
with taking the
Poisson bracket with the original Hamiltonian:
\be
\hat{O} \equiv \{ \frac{p^2}{2m} + V, ~~ \}~.
\ee
We have to note here that this treatment is entirely classical
and the operator $\hat{O}$ has nothing to do with
the quantum theory.
The solution for equation (\ref{nagy}) can be written
formally as:
\be
y = e^{x^0 \hat{O}} x~.  \label{y} 
\ee
The explicit form of this solution can be very complicated
and it depends on the form of the
potential $V$.
The other BRST-invariant variable can be ``built'' on $p$
and it is:
\be
\p =  e^{x^0 \hat{O}} p~. \label{pi}
\ee
These two variables $y$ and $\p$ are canonically conjugated to 
each other, as proved in the Appendix.
One can also prove that there are no more independent BRST-invariant
variables in the theory.

Two simple examples might give us a better feeling about these
variables.
If the potential is a constant, the two physical
variables are:
\be
y = x - \frac{p}{m}x^0~, \hskip 1cm \p = p~,
\ee
while for a potential of the form $
V = -A x$:
\be
y = x - \frac{p}{m}x^0 + \frac{A}{m} \frac{(x^0)^2}{2}~,
\hskip 5mm \p = p - Ax^0
\ee 
It is particularly simple to understand the first example.
The physical variable there, is $y$ which can be understood as
$x$ which evolves in $x^0$ as:
\be
\frac{dx}{dx^0} = \frac{p}{m}
\ee
and the momentum $\p = p$ conjugated to it.
The momentum is conserved in $x^0$.
This corresponds entirely to
what we expect to be physical. The second example gives us the
variables of a particle accelerated by a constant force.

One can also understand this result as follows:
the physical variables ($y$ and $\p$) in the extended phase space
correspond to variables $x$ and $p$ in the original
phase space, that evolve in time as:
\be
\frac{dx}{dt} = - e^{- x^0 \hat{O}} 
\part_t f(t) e^{x^0 \hat{O}}
\hat{O}x~,
\ee 
when $H$ does not depend explicitly on $t$.
For the case where $f$ only depends on $t$, the time-dependence
of $x$ becomes: $\frac{dx}{dt} = - \part_t f(t)
\hat{O}x~$.

For these variables $y$ and $\p$ to be physical they  have to be
co-BRST invariant as well. 
This demand gives a condition on the possible
gauge fixing conditions $f$.
\be
\{y , \Q\} =0 \Rightarrow % \left( \frac{\a p}{\m} + \b \right) \y =0
- \part_x y \part_p f + \part_p y \part_x f 
+ \part_0 y \part_{p_0}f = 0 
 \label{cond12}
\ee
By using (\ref{nagy}) we obtain:
\be
\part_x y [ -\part_p f + \frac{p}{m} \part_{p_0} f ]
+ \part_p y [ \part_x f - V' \part_{p_0} f] =0
\ee
The solution of this equation can be found if we decompose it as
follows:
\be 
-\part_p f + \frac{p}{m} \part_{p_0} f = \a(x,p,p_0) \part_p y  \label{f1}
\ee
\be
\part_x f - V' \part_{p_0} f = - \a(x,p,p_0) \part_x y
\ee
where $\a(x,p,p_0)$ can be any function of these variables.
We know that for all the possible potentials $ \part_p y =
- \frac{x^0}{m} + ... $. From the structure of the gauge fixing
condition it follows that $f$ does not depend on $x^0$.
This means that equation (\ref{f1}) can only be satisfied if
$\a$ vanishes.
This in turn means that the only $x,p,p_0$ dependence of the
gauge fixing function $f$ is through the constraint:
\be
f(t;x,p,p_0) = f(t;\f_2).            \label{f}
\ee
This result implies that there can be no further physical information
hidden in the gauge fixing conditions
related to the variables of the theory and  we can set
the gauge fixing condition to:
\be
\psi_2 = x^0 - f(t)~.                \label{fegysz}
\ee
The implication of (\ref{f}) is that the constraint algebra
simplifies and (\ref{kenyszalg4}) becomes:
\be
\{ \v_2 , \psi_2 \} =   -1~,     \label{kenyszalg4uj}
\ee
simplifying the t-evolution of all the variables 
(\ref{ido1})-(\ref{ido4}).
The time-evolution (along $t$) of the physical variables is
\be
\dot{y} = \dot{\p} = 0~,
\ee
which is not surprising since they by definition commute
with all the constraints and gauge fixing conditions.
%$t$ is thus only a parameter, along which there is
%no physical evolution.

Under these conditions one can  find an effective Hamiltonian
that generates the same t-evolution for all the variables
that we have here.
\be
\frac{dx}{dt} = \frac{p}{m} \part_t f = \{x, H_{eff} \} = \part_p
H_{eff} \label{hef1}
\ee
\be
\frac{dp}{dt} = -\part_x V \part_t f = \{p, H_{eff} \} 
= - \part_x H_{eff} \label{hef2}
\ee
\be
\frac{dx^0}{dt} = \part_t f \part_{p_0} H_{eff} \label{hef3}
\ee
It is important to note again, that unlike in other  cases of systems
with 
time dependent constraints, in the case
of reparametrization invariant models, the effective Hamiltonian is
not the physical Hamiltonian. 
The effective Hamiltonian here is defined on the whole
phase space, while the physical one will be only defined
on the physical phase space.

The conditions for the existence of $H_{eff}$ are
fulfilled:
\begin{eqnarray}
\part_p \part_x H_{eff} = \part_x \part_p H_{eff} 
& \Leftrightarrow &
 \part_p \left(
\part_x V \part_t f \right) =
\part_x \left(\frac{p}{m} \part_t f \right) = 
\frac{p}{m} \part_t \part_{\v_2} f~,
\\
\part_p \part_{p_0} H_{eff} = \part_{p_0} \part_p H_{eff}
& \Leftrightarrow &
\part_{p_0} \left( \frac{p}{m} \part_t f \right)
= \part_p \part_t f = \frac{p}{m} \part_t \part_{\v_2} f~,
\\
\part_x \part_{p_0} H_{eff} = \part_{p_0} \part_x H_{eff}
& \Leftrightarrow &
\part_x \part_t f = \part_{p_0} 
\left( \part_x V \part_t f \right) =
\part_x V \part_t \part_{\v_2} f~.
\end{eqnarray}
In other words, the conditions for the existence of
the effective Hamiltonian are the same as the conditions
for the existence of a consistent co-BRST charge.
>From the equations (\ref{hef1})-(\ref{hef3})
one can find the Hamiltonian by integration.
In the particular case when $f$ depends only on $t$
(\ref{fegysz}) the effective Hamiltonian is:
\be
H_{eff} = 
\left( p_0 + \frac{p^2}{2m} + V \right) \part_t f(t)~.  \label{heff}
\ee
The question is now whether there exists any preferred choice
for the time variable of the non-reparametrization invariant
theory. Is there anything that would point out the original
$x^0$ as an observable time? 
In some sense the answer is yes. 
One can interpret $x^0$ as the physical time of the system.
In that case the phase space is reduces to $x$ and $p$,
since $x^0$ becomes a parameter and $p_0$ looses any
physical meaning.
The physical
variables $y$ and $\p$ in the extended phase space can
be interpreted as the 
ones in the reduced phase space, which evolve along $x^0$ as
\be
\frac{dx}{dx^0} = \frac{p}{m} \hskip 1cm 
\frac{dp}{dx^0} = -\part_x V(x).               \label{eqm1}
\ee
This time evolution is independent (as expected) of the
gauge fixing, as long as the gauge fixing condition
respects the restrictions above. The physical Hamiltonian
is now the original one. There is a problem however, with this
interpretation: it can be valid only for a restricted
class of theories, where there exists the externally defined time,
thus for theories, where reparametrization invariance is
not a generic property.

This is why one can make a quite different interpretation,
namely, to consider $t$ as the physical time of the
system. In that case $H_{eff}$ is elevated to be the
physical Hamiltonian. 
The choice of the gauge fixing $f(t)$ has a physical meaning.
The choice $f(t) = t$ leads us back to the inertial
frame description and to equations of motion similar to
(\ref{eqm1}).

\setcounter{equation}{0}
\section{The Mass-less Relativistic Particle}

The most convenient action to start from in the
description of a relativistic particle is the one 
involving the ``ein-bein'':
\be
S = \int dt \frac{1}{2e(t)} \y_{\m\n} \frac{dx^{\m}}{dt}
\frac{dx^{\n}}{dt},
\ee
where the metric is $\y = diag(-1,1,1,1)$.
The momenta are conjugated to the $x^{\m}$ variables are:
\be
p_{\m} = \frac{\part L}{\part \dot{x}^{\m}} = \frac{x_{\m}}{e}.
\ee
The momentum conjugated to the $e$ variable
gives us the primary constraint:
\be
\v_1 = p_e = 0
\ee
The Hamiltonian becomes:
\be
H = \frac{e}{2} p_{\m}p^{\m} + p_e \l
\ee
and the consistency condition given by the
time evolution of the first constraint gives the expected:
\be
\v_2= \frac12 p_{\m} p^{\m}  \label{rel.kenysz}
\ee
Thus the BRST-charge is
\be
Q = \v_1 \ol{\cal{P}} + \v_2 \y.
\ee
The gauge fixing conditions can be written as
\be
\v_3 = e - f(t;x^i, p_{\m}) = 0
\ee
\be
\v_4 = x^0 - g(t;, x^i, p_{\m}) =0
\ee
The t-evolution given by the Gitman-Tyutin method
is consistent with these gauge fixing conditions,
and there are no restrictions upon them coming from
t-evolution. 
The physical variables are those that have
vanishing Poisson brackets with
both the BRST- and the co-BRST-charge.
Using the same method as in the previous 
Section one finds that the physical variables 
can be written as:
\be
y^j = e^{\frac{x^0 p_i \part^i}{p_0}} x^j = x^j + \frac{x^0}{p_0} p^j
\label{fizvalt.rel}
\ee
and their conjugated momenta
\be
\p_j =  e^{\frac{x^0 p_i \part^i}{p_0}} p_j = p_j. \label{fiz2p}
\ee
The demand, that the Poisson bracket between these
variables and the co-BRST should vanish, imposes
some restrictions on the possible form of the
gauge fixing conditions. 
Because of the simple form of $\Q$ we can treat the two
gauge-fixing functions separately.
\be
\{ \p_i, \Q \} = 0 \Rightarrow \part_i f(t;, x^j,p_j)
 = \part_i g(t; x^j,p_j) =0
\ee
thus none of the gauge fixing conditions depend on $x^{\m}$.
This in turn leads to:
\be
\{ y^i, f(t;p_{\m}) \} = \part_{p_i} f + 
\frac{p^i}{p_0} \part_{p_0} f =0
\ee
which means that $f$ can only depend on the momenta
through the second constraint:
\be
f(t; p_{\m}) = f(t;\v_2).
\ee
The situation with the second gauge-fixing condition is
somewhat more complicated, since there exist no 
possible functions $g(t;p_{\m})$ that would  
give $\v_4$ a  vanishing Poisson bracket with the expectedly
physical variable $y^i$. 
The only possibility is to find a gauge-fixing condition
whose Poisson bracket with $y^i$ is proportional to 
the gauge-fixing condition itself.
\be
\{ x^j + \frac{x^0}{p_0} p^j, x^0 - g(t; p_{\m}) \}
= \a(p_{\m}) (  x^0 - g(t; p_{\m}) ) \label{g}
\ee
where $\a(p_{\m})$ is a function. % to be found as follows.
%The terms in (\ref{g}) proportional to
%$x^0$ should vanish by themselves. This gives:
%\be
%\a(p_{\m}) = \frac{p^j}{(p_0)^2}
%\ee
%and this gives for the rest of the terms:
%\be
%p_0 \part_{p_j} g + p^j \part_{p_0} g = 
%\frac{p^j}{p_0} g.
%\ee
%If one wants to find the solution of this equation
%by expanding $g$ in terms of $p^{\m}$:
%\be
%g = g_a + g_{\m} p^{\m} + g_{\m \n}p^{\m}p^{\n} + ...
%\ee
%one finds only one solution, namely:
%\be
%g = g_0 p^0 
%\ee
%where $g_0$ can be any real number.  
There exists only one class of solutions to (\ref{g}):
$\v_4 = x^0 + g_0 p^0 t$.
A special case of this is the standard proper time
gauge %fixing condition
with $g_0 = \pm \frac{1}{m}$:
\be
\v_4 = x^0 \pm \frac{p^0}{m} t
\ee
We note here that this classical formulation is
not very sensitive to the introduction of a mass term
into the Lagrangian:
\be
\D L = - \frac{e}{2} m^2.
\ee
The secondary constraint (\ref{rel.kenysz}) changes to become:
\be
\v'_2 = \frac12 (p_{\m} p^{\m} + m^2)
\ee
but since the mass is a scalar, the Poisson bracket algebra
does not change. Neither do the physical variables.

The $t$-time evolution of the physical variables
vanishes and this means that the physical variables 
in the extended space can be understood as those variables
($x^i, p_i$) in the
physical subspace that evolve along $x^0$ as:
\be
\frac{dx^i}{dx^0} = \frac{p^i}{\sqrt{p^i p_i + m^2}}~, \hskip 5mm 
\frac{dp_i}{dx^0} =0~.
\ee
The physical Hamiltonian that would generate this evolution
is:
\be
H_{ph} = \sqrt{p^i p_i + m^2}
\ee 
We have come now to the important question: does the
BRST-co-BRST formalism distinguish a specific time
variable? Does it give information about which is the
physical time variable? The answer is again: yes, there
can exist an interpretation, where it does.
First of all, the physical variables as chosen 
in equation (\ref{fizvalt.rel}) could not have been
chosen in any other way. It is only the $p_0$
which is certain not to vanish. In the particle's
own reference frame all the other momenta are null,
but there exist no reference frame, where $p_0$
would vanish.
The same thing can also be said about the light-cone
coordinates: one can always find a frame, where $p_0 \pm p_1$
vanish.
This means that equation (\ref{fizvalt.rel}) is the unique 
choice of the physical configuration space variable.
>From here also follows that in this interpretation 
$x^0$ plays the role of the physical time.

For the reasons specified in the previous Section, 
the interpretation with more generality is the one, which
considers $t$ as the physical time of the system.
The effective Hamiltonian is:
\be
H_{eff} = g \frac{p_{\m} p^{\m}}{2} + p_e \part_t f~.
\ee
The freedom obtained this way (we can freely choose the
value of $g$) lets us include both the particles and the antiparticles
in the same formalism.

The BRST-co-BRST quantization of the relativistic particle
 model 
can be done as in the general case of Section 3.
The only thing  one has to  note is the form of the
physical creation and annihilation operators. These are of the
standard form $p_i \pm x_i$, and these operators commute
with both the BRST and the co-BRST operators.

\setcounter{equation}{0}
\section{Discussion}

Since there are many important reparametrization invariant theories,
it was interesting to see what can we learn about them
by the BRST-co-BRST method.
On the treated examples we could notice that the BRST-co-BRST
mechanism imposed some restrictions on the possible
gauge choices, though it did not give a definite answer 
on what gauge fixing conditions should be used,
 that
is: it did not show what the time variable should be.
In the two examples we treated,
the conditions for the existence of  a 
consistent and conserved co-BRST charge,
showed that the gauge-fixing of the original time variable ($x^0$) 
to the time-parameter ($t$) could only be done through
functions $f(t, \v_i)$, that is: functions depending on
the phase space variables only through the reparametrization
constraint. 

Time evolution in reparametrization invariant systems 
can be found in two ways. One way is to follow the
Gitman-Tyutin procedure, and define time evolution
of the physical variables directly through the
time-dependence of the gauge-fixing conditions.
The expressions one gets, are relatively simple, 
because the total Hamiltonian vanishes, thus all
time-evolution is defined by the evolution
of the constraint surface.
One can also define an effective Hamiltonian,
which generates the same evolution by Poisson brackets
(or commutators). 
This way one ``translates'' the time evolution of the
system to a more familiar language. 
The conditions for the existence of the
effective Hamiltonian were the same as the
conditions for the existence of the co-BRST charge.
This is one aspect of the relations between the
BRST-co-BRST scheme and time evolution.

By fixing reparametrization invariance, one fixes 
the  reference frame. 
For a detailed analysis on this aspect see \cite{f-g}.

The BRST-co-BRST quantization of 
reparametrization invariant systems
leads to well defined inner product states.
This, in concordance with earlier result also 
demanded that one introduces some imaginary eigenstates
to half of the operators.
We started with two classes of BRST-invariant states 
that were of the simple form of a matter state times a
ghost state. These states were neither inner product
states, nor co-BRST invariant. 
Then by a gauge regularisation we obtained the physical
states, that exhibited both these important properties.
We also noticed the $SL(2,R)$ algebra between the
regulators.

We only studied time-reparametrization invariant theories
and even between those, such theories that could be understood
as reparametrization invariant extensions of non-singular
theories. It is also possible to generalize the method
to theories with singular Lagrangians. This gives us
hope that the BRST-co-BRST method might give us some insight
even in the case of theories with several reparametrization
invariances, e.g. string theory or gravity.

%and the freedom for choosing the gauge fixing conditions
%lets us include both the particles and the antiparticles
%in the same formalism.

{\bf Acknowledgement} This work was supported by the FAPESP-grant
No 96/03108-0. I am very grateful to Prof. D.M. Gitman for
many useful discussions. 

%\bf{Appendix}

\setcounter{section}{1}
\setcounter{equation}{0}
\renewcommand{\thesection}{\Alph{section}}%%%%%%%%%%%%%%%%%%%%%%%%%%%%%%%%%%%%
\newpage
\noindent
{\Large{\bf{Appendix A}}}
 \vspace{5mm}\\
{\bf The proof of $y$ and $\p$ being canonically conjugated}\\ \\

The two BRST-invariant variables in the case of the
non-relativistic particle are given by (\ref{y}) and (\ref{pi}):
\be
y = e^{x^0 \hat{O}} x  \label{y1}
\ee
\be
\p =  e^{x^0 \hat{O}} p \label{pi1}
\ee
We notice that 
\be
p = -m \hat{O} x.
\ee
Using this fact and the Jacobi identities
we find that
\be
\{ y, - \frac{\p}{m} \} = 
\sum^{\infty}_{n=0} (x^0)^n W_n \label{w}
\ee
where
\be
W_n = \sum^{[\frac{n}{2}]}_{j=0}
\frac{n+1-2j}{(n+1-j)! j!} \{ \hat{O}^j x, \hat{O}^{n+1-j} x \}
\ee
and $[\frac{n}{2}]$ is the integer part of $\frac{n}{2}$.
One can use the Jacobi identities to prove that
\be
\hat{O} \{ \hat{O}^j x, \hat{O}^i x \} = 
\{ \hat{O}^{j+1} x, \hat{O}^i x \} + \{ \hat{O}^j x, \hat{O}^{i+1} x \}
\ee
Using this relation we obtain:
\be
\hat{O} W_n = (n+1) W_{n+1}. \label{wn+1}
\ee
We prove this only for odd $n$ but the proof for even 
$n$ is very similar.
\be
W_n = \sum^{\frac{n-1}{2}}_{j=0}
\frac{n+1-2j}{(n+1-j)! j!} \{ \hat{O}^j x, \hat{O}^{n+1-j} x \}
\ee
\be
\hat{O} W_n = \sum^{\frac{n-1}{2}}_{j=0}
\frac{n+1-2j}{(n+1-j)! j!} \{ \hat{O}^{j+1} x, \hat{O}^{n+1-j} x \}
+ \sum^{\frac{n-1}{2}}_{j=0}
\frac{n+1-2j}{(n+1-j)! j!} \{ \hat{O}^j x, \hat{O}^{n+2-j} x \}
\ee
If in the first term we introduce a new index notation
\be
k = j+1,
\ee
we have
\be
\sum^{\frac{n+1}{2}}_{k=1}
\frac{n+1-2(k-1)}{(n+1-(k-1))! (k-1)!} \{ \hat{O}^{k} x, \hat{O}^{n+2-k} x \}
\ee
changing the notation back $k \rightarrow j$ the whole
sum becomes:
\[
\hat{O} W_n = 
\frac{n+1}{(n+1)! 0!} \{ \hat{O}^{0} x, \hat{O}^{n+2} x \}
+ \sum^{\frac{n-1}{2}}_{j=1}
\frac{(n+1)(n - 2j + 2)}{(n+2-j)! j!} \{ \hat{O}^j x, \hat{O}^{n+2-j}
x \}+ \]
\be
+\frac{n+1}{(n+2- \frac{n+1}{2})! (\frac{n+1}{2})!} 
\{ \hat{O}^{(\frac{n+1}{2}} x, \hat{O}^{n+2-\frac{n+1}{2}} x \}
\ee
which can be added together to obtain:
\be
\hat{O} W_n = (n+1)
\sum^{\frac{n+1}{2}}_{j=0}
\frac{n - 2j + 2}{(n+2-j)! j!} 
\{ \hat{O}^j x, \hat{O}^{n+2-j} x \} \equiv (n+1)W_{n+1},
\ee
This equation is exactly what we wanted to prove.
It means that it is enough to find $W_0$ because the other
ones are of the form:
\be
W_n = \frac{\hat{O}^n}{n!} W_0
\ee
Since $W_0 = - \frac{1}{m}$ which is a constant, all the other
$W_n$ vanish. So from (\ref{w}) we have
\be
\{ y, \p \} = 1
\ee
thus they are canonically conjugated.

\end{document}